\documentclass[prb,amsmath,twocolumn,showkeys,showpacs]{revtex4}
\usepackage{graphicx}
\usepackage{dcolumn}
\usepackage{amssymb}
\usepackage{bm}
\usepackage{ulem}
\begin{document}
\newcommand{\be}{\begin{equation}}
\newcommand{\ee}{\end{equation}}
\newcommand{\vk}{\mathbf{k}}
\newcommand{\vq}{\mathbf{q}}
\newcommand{\vp}{\mathbf{p}}
\newcommand{\vecr}{\mathbf{r}}
\newcommand{\vx}{\mathbf{x}}
\newcommand{\caH}{\mathcal{H}}
\newcommand{\la}{\langle}
\newcommand{\ra}{\rangle}
\newcommand{\e}{\epsilon}
\newcommand{\half}{\frac{1}{2}}
\title{ X-ray edge singularity of bilayer graphene}
\author{Hyun C. Lee}
\email[email:~]{hyunlee@sogang.ac.kr}
\affiliation{Department of Physics and Basic Science Research Institute,
 Sogang University, Seoul, Korea}
\date{\today}
\begin{abstract}
The X-ray edge singularity of bilayer graphene is studied by generalizing the path integral approach 
based on local action which was employed for monolayer graphene. 
In sharp contrast to the case of monolayer graphene,  
the bilayer graphene is found to exhibit the edge singularity even at half-filling and 
its characteristics are determined by interlayer coupling.
At finite bias the singular behaviors sensitively depend  
on the relative magnitude of fermi energy and applied bias, which is due to the peculiar shape of energy band 
at finite bias.
\end{abstract}
\maketitle
\uline{Introduction}-
Physics of graphene systems is very rich.\cite{graphene_review}
A monolayer graphene has  massless Dirac fermion band whose density of states at Fermi energy vanishes at 
half-filling, and this is responsible for many of its semimetal properties.\cite{graphene_review}
A bilayer graphene is composed of two monolayer graphenes connected  by interlayer coupling 
in Bernal stacking structure.\cite{Falko,geim2006,Bernal}
The band structure of a bilayer graphene can be controlled externally by applying a gate bias between two layers.\cite{bias}
The (low energy) band of unbiased bilayer graphene is that of massive Dirac fermion \cite{Falko} without gap between electron and hole bands,
while a bias opens energy gap between them.

Meanwhile the transport properties of graphene systems  are intensively studied,
the optical studies also  revealed many interesting physical properties.\cite{review}
Among diverse optical probes we focus on the (near-edge) X-ray absorption spectroscopy \cite{book} and 
X-ray photoemission. 
The physics of  X-ray edge problem of \textit{fermi liquids} is very fascinating and well understood.\cite{x-ray_review,mahan1,
mahan2,anderson,nozieres,schotte}
The incident high-energy X-ray photon excites a deep core electron, leaving behind a (positively charged) core hole
which can be treated to be immobile in many cases.\cite{x-ray_review,mahan1}
If the excited electron escapes into vacuum  this is called X-ray photoemission process, and if 
the electron cannot escape this is referred to as X-ray absorption process.
Now the fermi sea of conduction electrons forming fermi liquid reacts to this \textit{suddenly}  created 
(namely, time-dependent) potential
which can be assumed to be well localized for the energy in the vicinity of absorption edge.\cite{x-ray_review,mahan1}
The conduction electrons interact with the potential in two distinctive ways:
the excitonic processes\cite{mahan2} which is essentially attraction between hole and conduction electrons and 
the orthogonality catastrophe\cite{anderson} which means a vanishing overlap between the ground state wavefunctions of 
conduction electrons before and after the creation of the hole.
Both excitonic processes and  orthogonality catastrophe are singular near X-ray absorption edge for  fermi liquids 
in the sense that there are divergences
in the perturbative expansion, so they require nonperturbative treatments-that is solving a particular 
type of singular integral equation exactly.\cite{nozieres}
This singular feature is strongly dependent on the fermi liquid nature of conduction electrons, especially the finite
density of states at fermi energy.
Then it is natural to ask what will happen for a graphene  which has a semimetal character at half-filling.

Motivated by the above observation the author has studied the X-ray edge problem of monolayer graphene 
in ref.[\onlinecite{hyun}]  (hereafter called (I), and see also  ref.[\onlinecite{guinea07}]).
The results for the monolayer graphene can be summarized as:(1) at half-filling the edge singularity is eliminated due to the vanishing density of states at fermi energy
(2) away from half-filling  the  edge singularity revives, and the exponent of power law behavior of
 the singularity depends on fermi energy  non-monotonously.

In this Brief Report we report  the results of on the X-ray edge problem of \textit{bilayer} graphene
which are  obtained by  generalizing the approach of (I), 
highlighting the differences with the case of monolayer graphene.
The most important finding  is that 
the X-ray edge singularity persists  for the bilayer graphene \textit{even at half-filling}, 
and this is essentially due to the nontrivial  matrix element of  1-particle Green's function which  stems from the 
interlayer coupling.
Also for the case with finite bias  the edge singularity shows up, but
its characteristics very sensitively depends on the relative magnitude of fermi energy and applied bias. 
This feature is due to the peculiar shape of the energy band  with finite bias.
Since the mathematical framework of approach in this report is almost identical with that of (I), we avoid duplicating 
the essentially identical mathematical expressions, instead we refer readers  to (I) for more details. 

\uline{Setup}- \;The honeycomb lattice of monolayer graphene is spanned by the following basis vectors:
$\mathbf{a}_1 = \ell (1,0)$ and $\mathbf{a}_2 =\ell(-\frac{1}{2}, \frac{\sqrt{3}}{2})$,
where $\ell = 2.46 \mathrm{\AA}$ is the lattice spacing.
Each site of $A$ sublattice is connected with three $B$ sublattice sites (assuming a $A$ site is at (0,0)):
$\mathbf{r}_1 = \ell ( 0, \frac{1}{\sqrt{3}})$, $\mathbf{r}_2 = \ell (-\frac{1}{2},-\frac{1}{2 \sqrt{3}})$, 
and $\mathbf{r}_3 = \ell (\frac{1}{2},-\frac{1}{2 \sqrt{3}})$.
A bilayer graphene has  Bernal stacking structure \cite{Bernal}, such that  $A$ site of top layer(=layer $t$) is sitting 
directly on the top of $B$ site of bottom layer (=layer $b$). The unit cell of bilayer graphene contains four sites, and 
we choose the ordering of the sites  according to $(bA,bB,tA,tB)\equiv(1,2,3,4)$.\cite{comment1}
The tight-binding Hamiltonian in Slonczewski-Weiss-McClure parameterization\cite{Bernal} 
includes one intra-layer hopping $\gamma_0$ and
three inter-layer hoppings $\gamma_1,\gamma_3,\gamma_4$. The most important characteristics of energy bands are determined 
by $\gamma_0$ and $\gamma_1$,\cite{review} hence $\gamma_{3}$ and $\gamma_4$ will be ignored in this paper.
The  values of $\gamma_0$ and $ \gamma_1$ are 
$\gamma_0 \sim 3.2 \,\mathrm{eV}$, $\gamma_1 \sim 0.4 \,\mathrm{eV}$.\cite{review}
We also take  into account the difference in site energy $u$ between two layers which can be controlled 
externally by applying bias voltage.\cite{bias}
Then the tight-binding Hamiltonian of bilayer graphene in momentum space takes the following form (spin indices are suppressed):
\be
\label{tight}
\hat{H}_0 = \sum_\vk \, C^\dag_\vk 
\left [ \begin{matrix}    u/2 &   \gamma_0 \Lambda_\vk &  0 &  0 \cr
     \gamma_0 \Lambda^*_\vk   & u/2 & \gamma_1 & 0 \cr
0  & \gamma_1 & -u/2 &  \gamma_0 \Lambda_\vk  \cr 
0 & 0 & \gamma_0 \Lambda^*_\vk  & -u/2 
        \end{matrix} \right ] C_\vk,
\ee
where $C_\vk = (c_{1,\vk},c_{2,\vk},c_{3,\vk},c_{4,\vk})^t$ ($t$ is matrix transpose) and
$c_{i,\vk}$ is the electron destruction operator with layer and sublattice index $i$. 
$\Lambda_\vk = \sum_{i=1}^3 e^{i \vk \cdot \mathbf{r}_i}$.
The diagonalization of eq.(\ref{tight}) gives four energy bands $\pm E_{\pm \vk} $, where
\be
E^2_{\pm, \vk} = \gamma_0^2 \vert \Lambda_\vk \vert^2 + \frac{\gamma_1^2}{2} + \frac{u^2}{4}
\pm \sqrt{ \frac{\gamma_1^4}{4} + (\gamma_1^2 + u^2) \gamma_0^2 \vert \Lambda_\vk \vert^2}.
\ee
Note that $ E_{+,\vk} \ge \gamma_1$.
$\pm E_{+, \vk}$ is the high energy electron (hole) band (often called dimer bands).
X-ray edge singularity is low energy process involving conduction electrons near fermi energy. Therefore,
we will be mostly interested in the low energy bands $\pm E_{-,\vk}$ which is gapless when $u=0$. To see this, 
note that $\Lambda_\vk$ vanishes at six corners of Brillouin zone, and among them only two are distinct:
$\mathbf{K}_+= ( \frac{4\pi}{3 \ell}, 0)$ and $\mathbf{K}_-= - \mathbf{K}_+$.
 This naturally introduces the valley index $\pm$.
Expanding $\Lambda_\vk$ around each valley $\mathbf{K}_\pm$, we obtain (for small $q_{x,y}$)
$\Lambda_{\mathbf{K}_+ + \vq} \sim \frac{ \sqrt{3} \ell }{2} ( - q_x + i q_y)$ and 
$\Lambda_{\mathbf{K}_- + \vq} \sim \frac{ \sqrt{3} \ell}{2} ( + q_x + i q_y)$.
Now in the small $\vq$  (continuum) limit, the Hamiltonian (\ref{tight})
 becomes 8x8 matrix ( 16x16 if spin included)
\be
\label{conti}
\hat{H}_0 = \int \frac{d^2 \vq}{(2\pi)^2} \, 
\left [ \begin{matrix} \Psi^\dag_+(\vq) \cr \Psi^\dag_-(\vq) \end{matrix} \right ]^t
\left [ \begin{matrix} \hat{K}_+ & 0 \cr 0 & \hat{K}_- \end{matrix} \right ]
\left [ \begin{matrix}  \Psi_+(\vq) \cr \Psi_-(\vq) \end{matrix} \right ],
\ee
where $\Psi(\vq) \equiv \sqrt{V} C_\vk$ ($V$ is the 
total volume of lattice) and $\Psi_{\pm}(\vq) = 
[\psi_{\pm 1}(\vq),\psi_{\pm 2}(\vq),\psi_{\pm 3}(\vq),\psi_{\pm 4}(\vq)]^t$.
$\hat{K}_+$ is the following 4x4 matrix for the valley $\mathbf{K}_+$:
\be
\label{Kmatrix}
\hat{K}_+ = \left ( \begin{matrix}
               \frac{u}{2}  & \pi(q) & 0 & 0 \cr
                 \pi^*(q) &  \frac{u}{2} & \gamma_1 & 0 \cr
0 & \gamma_1 & -\frac{u}{2} & \pi(q) \cr
0 & 0&  \pi^*(q) & -\frac{u}{2}
              \end{matrix}  \right ),
\ee
where $v = \frac{\sqrt{3}}{2} \gamma_0 \ell /\hbar$ and 
$\pi(q) \equiv  v (-q_x + i q_y)$.
The matrix $\hat{K}_-$ for the valley $\mathbf{K}_-$ can be obtained through 
$\hat{K}_-= \hat{K}_+( q_x \to -q_x)$.

In this report we neglect high energy processes involving the dimer bands $\pm E_{+,\vq}$. We will take the 
$\gamma_1 \sim 0.4 \mathrm{eV}$ to be the highest energy scale of our problem, 
and the low momentum region is defined by the condition
$ v q \ll \gamma_1$ ($q = \sqrt{q_x^2+q_y^2}$).
It is also fairly reasonable to assume that $
\gamma_1 \gg |u|$. In realistic experimental conditions, the fermi energy $\mu$ is also much smaller  than $\gamma_1$.
If $u=0$ then  $  E_{-, \vq} \sim v^2 q^2 /\gamma_1$ in the low momentum region, which is the energy dispersion for
the massive Dirac fermions.
If $u \neq 0$  the low energy electron band has a local maximum at $q=0$ with $E_{-, \vq=0} = \frac{\vert u \vert}{2}$ 
and  a global minimum at $q_{\rm min} = \frac{\vert u \vert }{2 v} \sqrt{\frac{ 2 \gamma_1^2 + u^2}{\gamma_1^2 + u^2}}$
with $E_{-, q_{\rm min}} = \frac{\vert u \vert }{2} \frac{ \gamma_1}{\sqrt{\gamma_1^2 + u^2}}$.
Since $ \vert u \vert \ll \gamma_1$, the above implies that the low energy band is  almost flat in low momentum region.
For conduction electrons to be available (electron doping assumed) with finite $u$, the fermi energy should be highter than 
$E_{-,q_{\rm min}}$.  We will assume $ \mu \ge |u|/2$.

The Hamiltonian of the (almost immobile) deep core electron is taken to be $ E_d d^\dag d$, where 
$E_d < 0$ is the core level energy ($d$ is the destruction operator of deep core electron).
The scattering potential by deep core hole is assumed to very local\cite{mahan1}, 
so that it is diagonal in layer and sublattice indices.
Then the interaction Hamiltonian between conduction electrons and deep core electron is given by
\be
\hat{H}_{\mathrm{int}}
=\int d^2 \vec{x} \,\Psi^\dag(\vec{x}) \delta(\vec{x}) \tilde{V} \Psi(\vec{x}) d^\dag d,
\ee
where $\tilde{V}$ is a 8x8 matrix representing intra-valley ($V_0$) and inter-valley ($V_1$) scattering:\cite{comment2}
\be
\label{potential}
\tilde{V} = V_0 \, \mathrm{I}_8 + V_1  \sigma_x \otimes \mathrm{I}_4,
\ee
where $\sigma_{x}$ is the 2x2 Pauli matrix acting on the valley space, and 
$V_1$ is assumed to be real for simplicity.
$\mathrm{I}_{4}$ and $\mathrm{I}_{8}$ is the identity matrix acting on layer-sublattice space and valley-layer-sublattice space,
respectivley.

\uline{Local action approach}- Since the scattering process occurs only at $\vec{x}=0$ our problem is essentially local, and it is advantageous 
to formulate the problem in a local way by integrating out the bulk degrees of freedom 
except for the one at $\vec{x}=0$.\cite{kane,nagaosa,hyun}
The resulting local action in imaginary time is (see (I) for more details)
\begin{align}
& S[\eta,d] = -\sum_{a,b} \int d \tau  d \tau' \bar{\eta}_a(\tau)
[G^{(0)}]^{-1}_{ab}(\tau-\tau') \eta_b(\tau')  \cr
&+ \int d \tau  \bar{d}(\tau) d (\tau)  [ \bar{\eta}(\tau) \tilde{V} \eta(\tau)] +
\int d \tau  \bar{d} ( \partial_\tau - \omega_T )d,
\end{align}
where $a,b$ are the valley-layer-sublattice indices and 
$\omega_T = \mu - E_d$ is the (bare) threshold energy. $\eta_a   = \Psi_{a}(\vec{x}=0)$ is the local
degrees of freedom.
$G^{(0)}_{ab}(\tau -\tau')$ is   the  local Green's function (zero temperature assumed hereafter):
\begin{align}
\label{localgreen1}
 G^{(0)}_{ab}(\tau -\tau') &\equiv -\la \Psi_{a}(\vec{x}=0,\tau) \Psi^\dag_{b}(\vec{x}=0,\tau') \ra \cr
&=\int \frac{d \epsilon d^2 \vq}{(2\pi)^3} e^{-i\epsilon(\tau -\tau')} G^{(0)}_{ab}(i\epsilon,\vq),
\end{align}
where 
\begin{align}
\label{localgreen2}
G^{(0)}_{ab}(i\epsilon,\vq) &=
 \left [ \begin{matrix} (i\epsilon+\mu) \mathrm{I}_4-\hat{K}_+ & 0 \cr
 0 & (i\epsilon+\mu) \mathrm{I}_4-\hat{K}_- \end{matrix} \right ]^{-1} \cr
& \equiv \left[ \begin{matrix} \hat{G}^+(i\epsilon,\vq) & 0 \cr 0 & \hat{G}^{-}(i\epsilon,\vq) \end{matrix} \right ].
\end{align}
The matrices $\hat{K}_\pm$ are defined in eq.(\ref{Kmatrix}) and 
$\hat{G}^\pm$ is the inverse matrix of $(i\epsilon+\mu) \mathrm{I}_4-\hat{K}_\pm$.
Upon angle interal of $\vq$, it is straightforward to check 
$ \int d \vq  \hat{G}^+ = \int d \vq  \hat{G}^-$.
The most essential difference between monolayer and bilayer graphene lies in the 
eq.(\ref{localgreen1}) as will be discussed in detail below.
The explicit momentum integral shows that only $\hat{G}^{\pm}_{ij}(\tau)$ with $(ij)=(11,44,22,33,23,32)$ 
are non-vanishing and that $\hat{G}^{\pm}_{23} = \hat{G}^{\pm}_{32}$.

\uline{ Nozi\`eres and De Dominicis (ND)'s solution} \cite{nozieres}-
Let us consider a \textit{single} species of fermion  whose (unperturbed) local Green's function in the long time asymptotic limit
is given by $G^{(0)}(\tau)= -\rho /\tau$, where $\rho$ is the density of states at fermi energy.
Then the Green's function in the presence of time-dependent potential 
$ -V \theta(\tau_1 -\tau) \theta(\tau -\tau_2) $ ($\theta(x)$ is step function and $V$ is a positive constant)
satisfies the following singular integral equation,
\be
\label{integral_equation}
G(\xi,\xi') = G^{(0)}(\xi-\xi') + \int^{\tau_1}_{\tau_2} d \tau \,
G^{(0)}(\xi -\tau)(-V)G(\tau,\xi').
\ee
The solution of eq.(\ref{integral_equation}) which  is asymptotically exact \textit{in the long time limit} is
\begin{align}
\label{ND}
G_{{\rm ND}}(\xi, \xi' \vert \tau_1,\tau_2) &= [\cos^2 \delta] G^{(0)} (\xi -\xi')  \left [ \frac{(\xi - \tau_2) ( \tau_1 - \xi') }{
(\tau_1 - \xi)(\xi' - \tau_2)} \right ]^{ \frac{\delta}{\pi}},
\end{align}
with the scattering phase shift given by
\be
\delta = \tan^{-1} [ \pi V \rho ].
\ee

\uline{Correlation functions}- In our case there are many species of fermions, so that the 
 eq.(\ref{integral_equation}) becomes a matrix equation:
$ G^{(0)} \to G^{(0)}_{ab}$ and $V \to \tilde{V}$ [see eqs.(\ref{potential},\ref{localgreen1})]. Also,
the exact solution cannot be obtained for $G^{(0)}_{ab}(\tau)$ of general form.

The photoemission is related to core hole (Matsubara) Green's function [see eqs.(28,29) of (I)]
\begin{align}
\label{cor1}
D(\tau_1,\tau_2) &= \langle T_\tau \, d^\dag(\tau_1) d(\tau_2) \rangle \cr
&= \exp\Big[ \mathrm{Tr}\ln (1 - G^{(0)} \hat{V} ) \Big ],
\end{align}
where $\hat{V}$ is  the time-dependent potential (matrix)
\be
\hat{V}(\tau ) = - \tilde{V} \theta( \tau_1 -\tau) \theta(\tau -\tau_2).
\ee
The X-ray absorption is related to the hole-conduction electron response function [see eqs.(38,39) of (I) ]
\begin{align}
\label{cor2}
F(\tau_1,\tau_2) &= \sum_{a,b} \, \la T_\tau \,d^\dag(\tau_1) \eta_{a}(\tau_1)
\bar{\eta}_{b}(\tau_2) d (\tau_2 ) \ra \cr
&=-\theta(\tau_1 -\tau_2) D(\tau_1,\tau_2)  \sum_{a,b} G_{ab} (\tau_1 -\tau_2),
\end{align}
where $ G_{ab} (\tau_1 -\tau_2)$ is the solution of eq.(\ref{integral_equation}) in \textit{matrix} form.
Note that $\xi \to \tau_1 -\tau_c$ and $\xi' \to \tau_2 + \tau_c$ should be taken
to obtain $G_{ab}$ in eq.(\ref{cor2}) ($\tau_c$ is a short time 
cutoff).
The core hole Green's function $D(\tau_1,\tau_2)$ can be also obtained from $ G_{ab} (\tau_1 -\tau_2)$ 
by parameteric integral.\cite{nozieres,bosobook}
Therefore, the task is (1) to compute the unperturbed local Green's functions of eq.(\ref{localgreen1});
(2) substitute them into the matrix version of eq.(\ref{integral_equation}) and find its solution (if possible);
(3) do parameteric integral  to find $D(\tau_1,\tau_2)$, and finally obtain $F(\tau_1,\tau_2)$.
For the computation of  unperturbed local Green's functions of eq.(\ref{localgreen1})
it is convenient to consider two cases, $u=0$ and $u \neq 0$, separately.

\uline{Results for the unbiased case} - 
For $u=0$, $\hat{G}^{\pm}_{11}=\hat{G}^{\pm}_{44}$, $\hat{G}^{\pm}_{22}=\hat{G}^{\pm}_{33}$ hold.
In the long time limit we obtain
\begin{align}
\label{11}
\hat{G}^{\pm}_{11}(\tau) = \begin{cases}
                      -\frac{(\gamma_1 -\mu)}{8\pi v^2 \tau} + \frac{1}{8\pi v^2 \tau^2} & \; \tau > 0, \cr
-\frac{(\gamma_1 -\mu)}{8\pi v^2 \tau}
+\frac{ 1 - 2 e^{\mu \tau}}{8\pi v^2 \tau^2} & \; \tau <0.
                     \end{cases}
\end{align}
\begin{align}
\label{22}
\hat{G}^{\pm}_{22}(\tau) = \begin{cases}
                      -\frac{\mu}{8\pi v^2 \tau} -\frac{1}{8\pi v^2 \tau^2} & \; \tau > 0, \cr
-\frac{\mu}{8\pi v^2 \tau}
-\frac{ 1 - 2 e^{\mu \tau}}{8\pi v^2 \tau^2} & \; \tau <0.
                     \end{cases}
\end{align}
$\hat{G}^{\pm}_{23}(\tau)$ is identical with $\hat{G}^{\pm}_{22}(\tau)$ up to sign.
Eqs.(\ref{11},\ref{22}) should be compared with those of monolayer graphene.
\be
G_{\rm mono} (\tau > 0) =\begin{cases}
                      -\frac{\mu}{4\pi v^2 \tau} -\frac{1}{4\pi v^2 \tau^2} & \; \tau > 0, \cr
-\frac{\mu}{4\pi v^2 \tau}
-\frac{ 1 -  e^{\mu \tau}}{4\pi v^2 \tau^2} & \; \tau <0.
                     \end{cases}
\ee
It is clear $\hat{G}^{\pm}_{22}(\tau)$ is essentially identical with $G_{\rm mono} (\tau )$ up to a factor of 2
(which comes from neglected dimer band of bilayer).
However, $\hat{G}^{\pm}_{11}(\tau)$ is qualitatively different from $G_{\rm mono} (\tau )$.
The difference is most marked at half-filling $\mu=0$.
In this case, the leading term of $\hat{G}^{\pm}_{11}(\tau)$ is proportional to $1/\tau$, while 
that of $G_{\rm mono}$ is proportional to $1/\tau^2$.
It is easy to check that when the unperturbed local Green function is proportional to $1/\tau^2$ there is no edge 
singularities in perturbative expansion [see eq. (46) of (I)].
Therefore, we conclude that the X-ray edge singularity persists for the bilayer graphene \textit{even at 
half-filling}.
Next let us compare $\hat{G}^{\pm}_{11}(\tau)$ with $\hat{G}^{\pm}_{22}(\tau)$.
The difference at half-filling $\mu=0$ is manifest, and we also note that even away from half-filling ($\mu \neq 0$)
$\gamma_1 \gg \mu$ holds, so that the most dominant contribution comes from the 
$-\gamma_1 /8 \pi v^2 \tau$ piece of  $\hat{G}^{\pm}_{11}(\tau)$ (and $\hat{G}^{\pm}_{44}(\tau)$) which is 
the premise for the ND solution.
This appearance of $\gamma_1$  is due to the non-trivial matrix element of 1-electron Green's function
$\la 0 \vert \psi_{11}(\vec{x}=0) \vert \vq \ra$.
Then the matrix integral equation is almost diagonal except for the potential matrix $\tilde{V}$.
However, this can be dealt with the ansatz [ see eqs.(20,21) of (I), there $\tau_x$ should be corrected to 
$\mathrm{I}_2$ ],
\be
G(\tau)  =\mathrm{I}_2 \otimes G'(\tau) + \sigma_x \otimes G^{\prime \prime}(\tau),
\ee
where $G'$ and $G^{\prime \prime}$ are the diagonal 4x4 matrices whose [22] and [33] entries are  absent, and 
[11] and [44] entries  are identical.
Then the solution of integral equation is given by[see eq.(25) of (I)] 
\begin{align}
 G^{\prime,\prime \prime}_{11,44}(\tau > 0) &\sim \frac{1}{2} \frac{(-\gamma_1) \cos^2 \delta_+}{\tau}\,
\left( \frac{\tau}{\tau_c} \right )^{2 \delta_+/\pi} \cr
&\pm \frac{1}{2} \frac{(-\gamma_1) \cos^2 \delta_-}{\tau}\,
\left( \frac{\tau}{\tau_c} \right )^{2 \delta_-/\pi},
\end{align}
where the scattering phase shift $\delta_\pm$ is given by
\be
\delta_\pm =\tan^{-1} [ \pi (V_0 \pm V_1 ) \frac{\gamma_1}{8\pi v^2}].
\ee
Note that $ \frac{\gamma_1}{8\pi v^2}$ is the counterpart of the density of states at fermi energy for fermi liquids.
The core-hole Green's function is given by 
\be
\label{result1}
D(\tau > 0 )  \sim e^{-\omega_T^* \tau} \, \frac{1}{(\tau/\tau_c)^{N_c (\delta /\pi)^2}},
\ee
where $N_c = 4 $ ( ${}^\pm 11, {}^\pm 44$ ) and $\delta^2 = (\delta_+^2 + \delta_-^2)/2$.
$\omega_T^*$ is the renormalized threshold energy.
Finally the X-ray absorption intensity which can be obtained from $F(\tau)$ by Fourier transform and analytic 
continuation is given by ($E_c$ is energy cutoff)
\be
\label{result2}
I(\omega) \sim \theta(\omega -\omega_T^*)\, \frac{\gamma_1}{8\pi v^2} \left( \frac{\omega - \omega_T^*}{E_c} \right )^{-
2 \delta_+/\pi + N_c (\delta /\pi)^2}.
\ee

\uline{Results for the case with finite bias}-
At finite $u$, asymmetry between top and bottom layer is expected, and this is reflected in the 
structure of local Green's function.
In this case the unperturbed local Green function is determined by the momentum region around
fermi line. Recalling the assumption $\mu \ge u/2$, we find that
the fermi momentum is given by
\be
q_0^2 = \frac{1}{v^2} \, \Big [ \mu^2 +u^2/4 + \sqrt{ u^2 \mu^2 + \gamma_1^2 (\mu^2 -(u/2)^2)}\Big ].
\ee
Then have $
E_{-,\vq} \approx  v^2 (q^2 - q_0^2) \frac{c}{\mu},$
where $c$ is a dimensionless constant given by
\be
\label{c}
c = 
 \frac{\mu^2 -(\frac{u}{2})^2}{2\gamma_1^2} + \sqrt{ \frac{ u^2 \mu^2}{4\gamma_1^4} + 
\frac{ (\mu^2 -(\frac{u}{2})^2)}{4\gamma_1^2}} +
\frac{u^2 v^2 q^2_0}{4 \gamma_1^4},
\ee
which \textit{becomes very small in the limit} $ \mu \to u/2$. This feature originates from almost  flat band 
strucutre at finite $u$ which was mentioned previously.
The long time limit of unperturbed Green's functions can be computed using the 
Laplace method of asymptotic analysis:
\be
 G^{\pm}_{ij}(\tau ) = -\frac{1}{c}\, \frac{ \mu}{8\pi v^2 \tau}  z_{ij},
\ee
where $z_{ij}$ are dimensionless constants whose detailed forms do not concern us here.
The constants just show that  $G^{\pm}_{22,33,23}$ can be ignored compared to the $G^{\pm}_{11,44}$. When $\mu$ is close to $u/2$,
 only $G^\pm_{11}$ gives a dominant  contributions.
In this case, the results eqs.(\ref{result1},\ref{result2}) still apply with the modification 
\be
\gamma_1 \to \frac{\mu z_{11}}{c}, \quad N_c =2,  \;\;  ( {}^\pm 11).
\ee
Owing to the factor $c$ of eq.(\ref{c}), a very interesting variation of correlation function as a function of 
fermi energy and bias is expected.

\uline{Remarks and  summary}-The 
near edge X-ray absorption experiment for graphene systems was reported in  refs.[\onlinecite{experiment1}].
The K edge singularity around 283 eV is clearly visible. However, the fermi energy and the bias dependence, which 
is our main point, was not studied 
in this experiment. For more precise comparison, we have to include the effects of  the (ignored) dimer band, 
the contribution from C 1$s$ $\sigma^*$ transition, the band dispersion of core level, and various broadening effects 
such as lifetime and temperature, and full treatment of these are beyond the scope of this Brief Report.

In summary, we have studied the X-ray edge singularity of bilayer graphene by generalizing the approach 
employed for monolayer graphene. The bilayer graphene exhibits the edge singularity even at half-filling.
Also at finite bias, the singular behaviors are found to depend very sensitively 
on the relative magnitude of fermi energy and applied bias.

\begin{acknowledgements}
The author is grateful to H. Cheong and H. Kim for very useful discussions.
This work was  supported by Mid-career Researcher Program through NRF grant funded by 
the Mest (No. 2010-0000179) and by the Special Research Grant of
Sogang University.
\end{acknowledgements}


\begin{thebibliography}{}
\bibitem{comment1} Note that our ordering is different from that of Ref.[\onlinecite{Falko}].
\bibitem{comment2} We point out a typo error in eq.(5) of (I): 
$\sigma_x \otimes \tau_x$ should read as $\sigma_x \otimes \mathrm{I}_2$.
\end{thebibliography}
\end{document}